\documentclass{acm_proc_article-sp}
\usepackage{balance}
\usepackage{indentfirst}
\begin{document}

\title{Generating ordered list of Recommended Items: \\a Hybrid Recommender System of Microblog}
\balance
\setlength{\parindent}{1em}
%
%
%
%
%

\numberofauthors{2} 
%
\author{
%
%
\alignauthor Yingzhen Li\titlenote{To whom correspondence should be addressed}\\
       \affaddr{Sun Yat-Sen University}\\
       \affaddr{135 Xingangxi Road}\\
       \affaddr{Guangzhou, China}\\
       \email{liyzhen2@mail2.sysu.edu.cn}
\alignauthor Ye Zhang\\
       \affaddr{Sun Yat-Sen University}\\
       \affaddr{135 Xingangxi Road}\\
       \affaddr{Guangzhou, China}\\
       \email{zhangye5@mail2.sysu.edu.cn}
}

\maketitle

\begin{abstract}
Precise recommendation of followers helps in improving the user experience and maintaining the prosperity of twitter and microblog platforms. In this paper, we design a hybrid recommender system of microblog as a solution of KDD Cup 2012, track 1 task, which requires predicting users a user might follow in Tencent Microblog. We describe the background of the problem and present the algorithm consisting of keyword analysis, user taxonomy, (potential)interests extraction and item recommendation. Experimental result shows the high performance of our algorithm. Some possible improvements are discussed, which leads to further study.
\end{abstract}


\terms{Algorithm, Machine Learning, Data Mining}


\section{Introduction}

Online social networking services like Twitter have been tremendously popular, with a considerable speed of user growth. Thousands of new registrations are observed everyday in dominant platforms like Sina and Tencent Microblog since the introduction of microblog - Chinese twitter - in 2007.
Celebrities and organizations also register microblog, which leads to diversity of topics and helps attract more potential users. However, flooded information can puzzle the users and even result in the loss of them. So reducing the risk of puzzlement and recommending attractive items - specific users selected for recommendation - are crucial for user experience improvement and prosperity maintenance, which present opportunities for novel machine learning and data mining approaches.

Recommender systems can be categorized into content-based algorithm \cite{Mc:Birds}, collaborative filtering \cite{Kon:apply}, and influential ranking algorithm \cite{Wang:Graph}.
Unfortunately, all of them consider little of user profile's fidelity, preference variance and interactions, causing difficulty of precise and stable recommendation. To overcome these weaknesses of single method, we construct a hybrid recommender system specified to Tencent Microblog, which generates ordered item list by mining the data of the platform \cite{Niu:Tencent}.

The rest of the paper is organized as follows. Section 2 will discuss the background of the problem, and Section 3 will describe the design of the hybrid recommender system, including keyword analysis, user taxonomy, preference extraction, discovery of potential interests and generation of ordered recommendation. Section 4 will present the training process. Section 5 will show the experimental results and discuss some improvements, and the paper will be concluded in Section 6.

\section{Background}
Observing the popularity of twitter services, Tencent, one of China's leading Internet service portal, launched its microblog platform - Tencent Microblog - in 2010. It has attracted a lot of registered users(425 million registered accounts and 67 million daily active users in season 1, 2012 \cite{QQ:Finance}) and became one of the dominant microblog platforms in China based on the large user group of its instant messaging service QQ(711.7 million \cite{Tencent} on Sep 30, 2011).
Celebrities and organizations - items carefully categorized into hierarchies - are invited to register the platform, leading to a nice growth in the user group.
Furthermore, Tencent's microblog service is embedded in its other leading platforms like Shuoshuo(signature of user's QQ account), Qzone(blog platform), Pengyou.com(SNS service) and Weixin(mobile messenger), hence user can write or comment a message directly on the website of Tencent Microblog or via the third-party port and related platforms.

While Tencent has the largest microblog user group, Sina Microblog takes a commanding lead with 56.5\% of China's microblog market based on active users and 86.6\% based on browsing time over its competitors \cite{Kyle:Sina}. This fake prosperity of Tencent Microblog, which is far from the public perception, results from the existence of the fake users(explained in section 3.2), widely used spammer strategy \cite{Baike:Zombie} and the weird definition of active users. Tencent Microblog considers those who frequently write(retweet or comment) or read microblog messages - no matter on the website or other associated platforms - as active users, while Twitter and Sina Microblog define them as those who login the platform everyday.
User messages generated via from other related platforms confuse the recommender finding the real interests of users, which leads to the decrease of acceptance.
Tencent Microblog users accept the recommendations in a low percentage(less than 9\% according to our survey \cite{Niu:Tencent}), and the recommended item lists isn't updated in time, which deviate from the users' present preferences.

\section{Algorithms}
As shown in the prior study, each of the existing algorithms has its unavoidable disabilities.
This section introduces a hybrid recommender system to overcome them, including the preparations - keyword analysis and user taxonomy - and the main part of it.

\subsection{Keyword Analysis}
Mining synonyms in user's keywords helps in finding their interests.
However, applying association rule algorithm \cite{Agrawal:AR} to find them directly in the huge keyword set is unrealistic since that involves searching all possible combinations. So we parallel this process by adopting revised FDM(Fast Distributed Mining of association rules) \cite{Han:FDM}, based on the downward-closure property of support which guarantees that the necessity and sufficient condition for a frequent itemset is the frequency of all its subsets \cite{Tan:DM}. Moreover, we insert the ambiguous keywords('apple' for instance) into different classes simultaneously, which cause the inconsistence of class size.

Let $U = \{u_1, u_2,..., u_m\}$ be the set of users where $u_i$ has the value of its ID. Each user $u_j$ has its keyword set $K_j = \{k_{j1}, k_{j2},..., k_{jn_j}\}$ with weights $\mathcal{W}_j = \{w_{j1}, w_{j2},..., w_{jn_j}\}$, and we denote $\mathcal{K} = \bigcup_{j = 1}^{m} K_j$ as the set of all users' keywords.
The database $\mathcal{DB}$(user-keyword set) is divided into n subsets $\mathcal{DB}_i$ and these subsets are broadcasted to the remote sites $\mathcal{RM}_i$. $\mathcal{T}^j$ is the result generated in the $j^{th}$ iteration:
$$\mathcal{T}^j = \{T^j_1, T^j_2,..., T^j_{n_j}\},$$
$$T^j_i = \{k^j_{i1}, k^j_{i2},..., k^j_{il_j}\},$$ where $k^j_{ip}$ are the keywords of the $i^{th}$ keyword transaction in the $j^{th}$ iteration. Apriori algorithm is applied to generate the candidate transaction set $\mathcal{C}^j_i$ at $\mathcal{RM}_i$ in the beginning of the $j^{th}$ iteration as follows:
$$\mathcal{C}^j_i = Apriori\_gen(T^{j - 1}_i),$$
where
$$\mathcal{C}^j_i = \{C^j_{i1}, C^j_{i2},...,C^j_{ih_j}\},$$
$$C^j_{il} = \{k^j_{il1}, k^j_{il2},..., k^j_{ilq_j}\}.$$
Let $\mathcal{A} = \{A_1, A_2,..., A_m\}$ be a set of mappings where
$$A_j: \mathcal{K} \rightarrow \mathcal{W}_j \cup \{0\},$$
\[
A_j(k_i) =
\begin{cases}
w_{jl}, & k_i = k_{jl} \in \mathcal{W}_j\\
0, & k_i \notin \mathcal{W}_j
\end{cases}.
\]
$\mathcal{RM}_i$ computes the local support and confidence of $C^j_{il}$ by
$$supp\_local(C^j_{il}) = average(\min A_p(k^j_{ilk}))$$
$$(k^j_{ilk} \in C^j_{il}, u_p \in \mathcal{DB}_i),$$
$$conf\_local(C^j_{il}) = \frac{supp\_local(C^j_{il})}{supp\_local(C^{j-1}_{il})},$$
and eliminates those which fail to satisfy local minimums $supp\_local$ or $conf\_local$. Then remote site sent the remaining candidate transactions $C^j_{il}$ to the polling site $\mathcal{PL}_K$, where $K = polling(C^j_{il})$ is a hash function.

$\mathcal{PL}_K$ gathers the candidate transactions $C^j_{il}$, computes global support $supp\_global(C^j_{il})$ and confidence $conf\_global(C^j_{il})$ by sending request to remote sites for local values' return:
$$supp\_global(C^j_{il}) = average(\min A_p(k^j_{ilk}))$$
$$(k^j_{ilk} \in C^j_{il}, u_p \in \mathcal{DB}),$$
$$conf\_global(C^j_{il}) = \frac{supp\_global(C^j_{il})}{supp\_global(C^{j-1}_{il})}.$$
Then $\mathcal{PL}_K$ filters out the candidates which fail to satisfy the constraint of $supp\_global$ and $conf\_global$. Generally the local minimums coincide with the globals. For convenience we still denote the updated candidate sets as $\mathcal{C}^j_i$.

Then home site gathers $\mathcal{C}^j_i$ from the polling sites to generate the result of transactions(keyword classes) in the $j^{th}$ iteration:
$$\mathcal{T}^j = \displaystyle \bigcup_{i} \mathcal{C}^j_i,$$
where
$$supp\_global(T^j_i) >= supp\_global,$$
$$conf\_global(T^j_i) >= conf\_global.$$
The process is terminated if no new transaction is generated, before termination the home site broadcasts the transactions $T^j_i$ to the remote site $\mathcal{RM}_K$ where $\mathcal{RM}_K$ is the original remote site of $T^j_i = C^j_{Km}$, and then starts next iteration.

The final result of keyword class is $$keyword\_class = \{class_1, class_2,..., class_N\},$$ where $$class_i = \{k_{i1}, k_{i2},..., k_{im}\}$$ is the set of synonyms.

The choice of minimums affects the precision and computational complexity tremendously. We sampled 1000 users' keywords and found out that these users have their keyword weights average in 0.14, so we assign $$supp\_local = supp\_global = 0.2,$$
a little higher than the average weight. $conf\_local$/$conf\_global$ is affected by $supp\_local$/$supp\_global$, in this case
$$conf\_local = conf\_global = \frac{0.14}{0.2} = 0.7.$$

\subsection{User Taxonomy}
\begin{figure}[htbp]

\centering

\includegraphics[width=8cm]{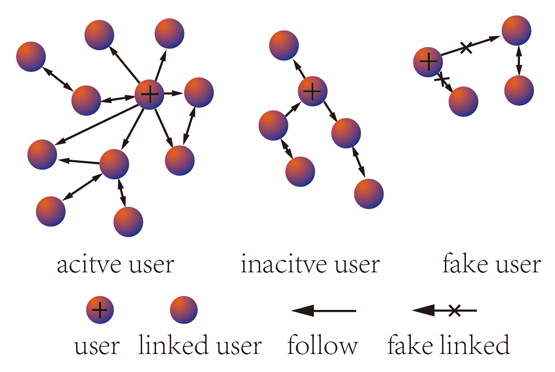}

\caption{User taxonomy. Active users have more followees than inactive users. Links related to fake users are eliminated since they're not real users in the user network.} \label{fig:taxonomy}

\end{figure}

Most of the microblog platforms divide their users into 2 groups - active and inactive - to apply different types of strategies. However, some users don't login Tencent Microblog directly while they have records of tweets generated from other related platforms, and these messages could hardly reflect their interests. Furthermore, they rarely interact with other users and have few favorites for the same reason. So we classify them as fake users(see Figure \ref{fig:taxonomy}). In addition, we also consider the spammers as fake since they seldom use microblog even indirectly.

Due to the absence of login records, the activeness function $act(u_j)$ counts the number of tweets and interactions and computes $u_j$'s activeness by applying the thresholds $min\_activeness$ and $min\_action$:
$$act(u_j) = tweet \times is\_fake(u_j),$$
where
\begin{equation*}
\begin{aligned}
is\_fake&(u_j)   \\
&= \frac{1+sgn(at+retweet+comment-min\_action)}{2}
\end{aligned}
\end{equation*}
and
\[
user\_class(u_j) =
\begin{cases}
\text{active}, & act(u_j) \ge min\_activeness\\
\text{inactive}, & 0 < act(u_j) < min\_activeness\\
\text{fake}, & act(u_j) = 0
\end{cases}.
\]
We assign $min\_activeness = 100$ and $min\_action = 20$ since only 33.2\% of the users have written more than 100 tweets(771599 in 2320895), and apply the algorithm to divide the user group into 3 classes.

An appropriate user taxonomy helps in improving the precision of recommendation. Users with similar favorites often accept similar items, hence dividing users into smaller groups by their interests can balance the precision and computational complexity. However, we haven't done this due to the sparsity of successful recommendation records, which reflect the user's interests directly.

\subsection{Generating Recommendations}
After keyword analysis and user taxonomy which are preparations of the recommendation, it comes the main part of our hybrid recommender system, consisting of item popularity ranking, (potential)interests discovery and the grading function, to generate recommended items and evaluate the possibility of acceptance or rejection. The system maps the users' (potential)interests to their corresponding item categories and grades selected candidates in these categories with indicators of similarity and popularity. It also contains special algorithms with respect to fake users in order to reach a precise recommendation.

\subsubsection{Item Popularity Ranking}
\begin{figure}[htbp]

\centering

\includegraphics[width=8cm]{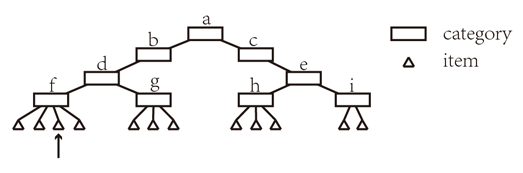}

\caption{Item categories organized in hierarchy. The pointed item belongs to the category a.b.d.f.} \label{fig:hierarchy}

\end{figure}

An item is a specific user, which can be a famous person, an organization, or a group. Items are organized in different categories of professional domains by Tencent to form a hierarchy(see Figure \ref{fig:hierarchy}). For example, an item, Dr. Kaifu LEE, is represented as \emph{science-and-technology.internet.mobile} \cite{KDD}.

The number of an item's followers indicates its popularity directly. Recommending hot items in a user's interested field promotes the possibility of acceptance effectively. For users who show little of their preferences(especially the fake users) we recommend the most popular items in the whole itemset.

Let $I = \{i_1, i_2,..., i_n\}$ be the item set and $h_k$ be the category(hierarchy) of an item $i_k$($h_k$ may coincide in some $h_j$, and the set of $h_k$ is denoted as $H$). Then the rank of $i_k$ in $h_k$ is computed by
$$hot_k = get\_hot\_rank(i_k, h_k)$$
where the function counts the number of $i_k$'s followers and return its ranking in $h_k$. Similarly the hot rank of $i_k$ in the whole item set $I$ is
$$HOT_k = GET\_HOT\_RANK(i_k).$$
To apply the rank information in our grading process, we normalize the rank by function
$$n(x) = \frac{1 + e}{1+e^x},$$
and for convenience we still denote the normalized results as $hot_k$ and $HOT_k$.

\subsubsection{Mining Interests from Keywords}
Users are inclined to accept items of their interests. Active users have more keywords which reflect their favorites, and we map these interests to the hierarchy $H$ to obtain candidate items.

Consider the keyword classes set
$$keyword\_class = \{class_1, class_2,..., class_N\}$$
generated by the keyword analysis. A mapping
$$\mathcal{KH}: H \rightarrow power(keyword\_class),$$
$$\mathcal{KH}(h_k) = \{class_{k1}, class_{k2},..., class_{kn_k}\}$$
is defined to construct the keyword class of a given category $h_k$(see section 3.3.4 for details).
Suppose a given user $u_j$(or a given item $i_k$) has keywords
$K_j = \{k_1, k_2,..., k_{n_j}\}$ with weights$\{w_1, w_2,..., w_{n_j}\}$ which satisfy $\sum_{i = 1}^{n_j} w_i = 1.$
A function
$$key\_class(u_j) = \{class_{j1}, class_{j2},..., class_{jm_j}\}$$
computes the keyword class of a given user $u_j$(or item $i_k$) where $class_{ji}$ satisfies
$$K_j \cap class_{ji}\ \neq \varnothing,$$
and the corresponding weight of $class_{ji}$ is
$$\overline{W}_{ji} = \sum_{k_l \in K_j \cap class_{ji}} w_l.$$
After keyword analysis of individuals the target categories $\{h_k\}$ is generated where $h_k$ satisfies
$$\mathcal{KH}(h_k) \cap key\_class(u_j) \neq \varnothing,$$ hence the candidate items are the items in each $h_k$(suppose $i_k$ included).
A vector function is defined on $U \cup I$ as follows:
$$class\_weight(u_j) = (weight_1, weight_2,..., weight_N)$$ (here $u_j$ can be substituted by $i_k$), where
\[
weight_i =
\begin{cases}
\overline{W}_{jl}, & class_i = class_{jl} \in key\_class(u_j)\\
0, & class_i \notin key\_class(u_j)
\end{cases}.
\]
The similarity between $i_k$ and $u_j$ is the normalized Euclid distance of these 2 vectors:
$$sim(u_j, i_k) = n(|class\_weight(u_j) - class\_weight(i_k)|)$$
where $n(x)$ is the normalization function mentioned in Section 3.3.1. For the sparsity of the non-zero values we don't compute these vectors directly, reducing the storage and computational complexity.

\subsubsection{Discovering Potential Interests}
\begin{figure}[htbp]

\centering

\includegraphics[width=8cm]{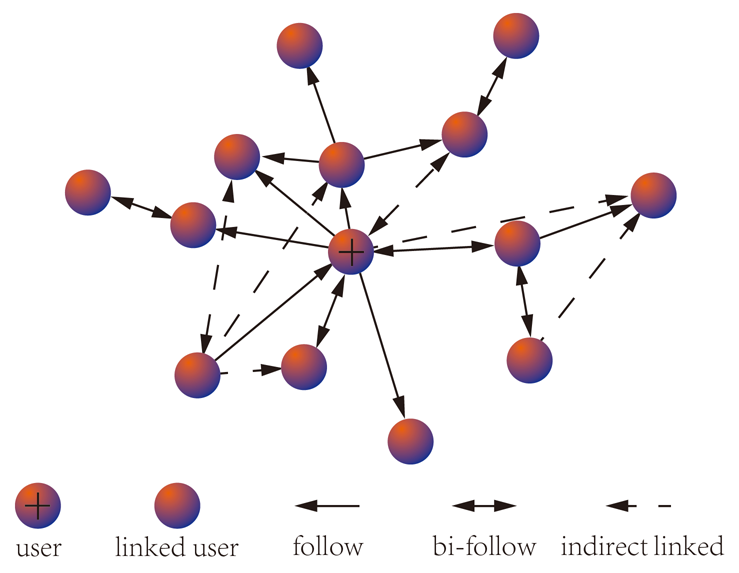}

\caption{User network with indirect links($depth = 2$). Indirectly linked users interact with each other even without followships. The length of the arrows represents the familiarity between two users(a user may be more familiar with some indirectly linked users than its followees).} \label{fig:familiarity}

\end{figure}

Few inactive users have enough keywords, hence we design indirect collaborative filter to mine their potential interests from their followees. We build up the social network of a user $u_j$ and search for its potential interests in its followees $followee(u_j) = \{u_k\}$ and even in its followees' followees(see Figure \ref{fig:familiarity}).

Let $depth$ be the maximal levels amount of the searching process, in fact $depth \leq 3$ is enough for the process to mine a user's potential interests. The related users of $u_j$ is
$$related\_users(u_j) = search\_followee(u_j, depth)$$
where $search\_followee(u_j, depth)$ returns the followees and indirect followees of $u_j$ with max level $depth$ in the social network. Then for every $u_k$ in $related\_users(u_j)$ we compute the keyword classes as mentioned above and merge them into the set
\begin{equation*}
\begin{aligned}
potential\_key(u_j) &= \{class_{j1}, class_{j2},..., class_{jn_j}\}   \\
&= \displaystyle\bigcup_{u_k \in related\_users(u_j)} key\_class(u_k)
\end{aligned},
\end{equation*}
where the $i^{th}$ keyword class $class_{ji}$ has the weight
$$\widetilde{W}_{ji} = \displaystyle\sum_{\substack{u_k \in related\_users(u_j), \\ class_{kl_k} = class_{ji}}} \overline{W}_{kl_k}fami(u_j, u_k).$$
$fami(u_j, u_k)$ computes the familiarity of $u_j$ and $u_k$ by adopting indicators of interactions(at(@), retweet and comment) which could only happen in linked users:
$$fami(u_j, u_k) = \omega_1f(at) + \omega_2f(retweet) + \omega_3f(comment),$$
where $\omega_i$ satisfying $\sum_{i = 1}^{3} \omega_i = 1, \omega_i \ge 0$ is obtained in training process and $f(x)$ is a sigmoid function
$$f(x) = \frac{2}{1+e^{-x}} - 1.$$
Finally we merge $key\_class(u_j)$ and $potential\_key(u_j)$ into the set
$$interests(u_j) = \{class_{j1}, class_{j2},..., class_{jn_j}\}$$
with weight of $class_{jl}$
\[
W_{jl} =
\begin{cases}
\overline{W}_{jl_1}, & class_{jl} \in key\_class(u_j)\\
\widetilde{W}_{jl_2}, & class_{jl} \in potential\_key(u_j)\\
\frac{1}{2}( \overline{W}_{jl_1} + \widetilde{W}_{jl_2} ), & class_{jl} \text{ in both sets}
\end{cases},
\]
where
$$class_{jl} = class_{jl_1} \in key\_class(u_j),$$
$$ class_{jl} = class_{jl_2} \in potential\_key(u_j).$$
Correspondingly the target category $\{h_k\}$ satisfies
$$\mathcal{KH}(h_k) \cap interests(u_j) \neq \varnothing,$$
and the candidate items are in each $h_k$ as mentioned before(suppose $i_k$ included).
We modify the vector function $class\_weight(u_j)$ which is defined in section 3.3.2 by using $interests(u_j)$ to substitute $key\_class(u_j)$, i.e. $W_{jl}$ instead of $\overline{W}_{jl}$, and compute the similarity of $u_j$ and $i_k$ as before. In this way, we get the similarity between items and inactive users. The algorithm can also be applied to the recommendation for the active users with smaller value of $depth$.

\subsubsection{Grading Function}
\begin{figure*}[htbp]

\centering

\includegraphics{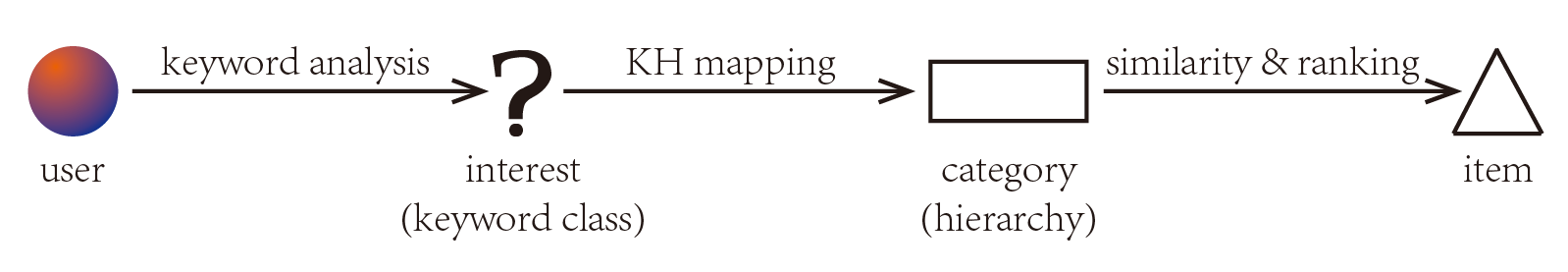}

\caption{Recommendation process. Keywords analysis extracts user's interests, $\mathcal{KH}$ mapping match these interests to corresponding item categories thus obtains the set of candidates, and the grading function assign grades of the recommendation by computing the similarity and popularity. } \label{fig:grading}

\end{figure*}

In studies above we presented the extraction of a given user $u_j$'s (potential)interests and the generation of the recommended candidates. The grading function $grade(u_j, i_k)$ computes the possibility of acceptance(positive grade) or rejection(negative grade) with indicators of $i_k$'s popularity and $sim(u_j, i_k)$ computed as above(see Figure \ref{fig:grading}). Then we pick out the first $k$ candidates and sort them in descending order to generate final recommendation, where in our case $k = 3$.

Let $I = \{i_1, i_2,...,i_n\}$ and $H = \{h_1, h_2,...,h_n\}$ be the item set and the category(hierarchy) set as previously defined and suppose we have extracted keyword classes of each user and item. We specify the definition of $\mathcal{KH}$(see section 3.3.2) as follows:
$$\mathcal{KH}: H \rightarrow power(keyword\_class),$$
$$\mathcal{KH}(h_k) = \{class_{k1}, class_{k2},..., class_{kn_k}\}$$
$\mathcal{KH}(h_k)$ computes the keyword classes of a given category $h_k$ with corresponding weight of $class_{kp}$
$$\hat{W}_{kp} = average(\overline{W}_{jl_j}),$$
$$i_j \in h_k, key\_class(i_j)(l_j) = class_{jl_j} = class_{kp}.$$
Here $power(keyword\_class)$ is the power set of $keyword\_class$.

We revise the definition of $class\_weight()$ by extending its domain to $H$:
$$class\_weight(h_k) = (weight_1, weight_2,..., weight_n),$$
\[
weight_i =
\begin{cases}
\hat{W}_{kp}, & class_i = class_{kp} \in \mathcal{KH}(h_k)\\
0, & class_i \notin \mathcal{KH}(h_k)
\end{cases}.
\]
As previously mentioned we don't compute these vectors directly.
The function $fond()$ is defined on $U \times H$ by
$$fond(u_j, h_k) = g(class\_weight(u_j) \cdot class\_weight(h_k), 100),$$
$$g(x, y) = \frac{2(1 + e^{-y})}{(1 - e^{-y})(1 + e^{-xy})} - \frac{1 + e^{-y}}{1 - e^{-y}}$$
to compute the ratio of $u_j$'s fondness for category $h_k$.

Finally the grading function of active/inactive users is
$$grade(u_j, i_k) = 2 fond(u_j, h_k) (\alpha_1hot_k + \alpha_2 sim(u_j, i_k)) - 1,$$
where $hot_k$, $sim(u_j, i_k)$ are indicators as mentioned above and $\alpha_i, i = 1, 2$ satisfying $\alpha_1 + \alpha_2 = 1, \alpha_i \ge 0$ are obtained in the training process. Valued in $[-1, 1]$, the grading shows the possibility of acceptance(positive grade) or rejection(negative grade). Considering the variance of user preferences in a certain period, a revised grading function is defined as
$$revised\_grade(u_j, i_k) = \frac{1}{\lambda} time(u_j, h_k) grade(u_j, i_k),$$
where
$$time(u_j, h_k) = 1+(\lambda - 1)e^t, t \in (-\infty, 0].$$
$t$ is the time of $u_j$'s latest acceptance of items in $h_k$(the current time is $t = 0$ and the default time is $t = -\infty$ if no acceptance). $\lambda$ indicates the proportion of recent interest(in our experiment $\lambda = 2$).

Fake users, especially those with few keywords, receive different treatment. Observing the difficulty to apply similarity and familiarity function, $potential\_key(u_j)$ is not generated, i.e. $interests(u_j) = key\_class(u_j)$, $W_{jl} = \overline{W}_{jl}$. The grading function of their recommendations only encompasses the hot rank and the preference to compute the grading
$$grade(u_j, i_k) = (1 + fond(u_j, h_k)) HOT_k - 1.$$
This definition emphasizes the popularity of the item in $I$ and increases the grading if $i_k$ and $u_j$'s interests coincide.

\section{Training}
The choice of parameters $\omega_i$ and $\alpha_i$ affects the performance of the hybrid recommender system we introduced above. The training process obtains sets of optimal parameters with respect to each user group by applying the supervised online training algorithm(convergence granted and time-saving).
Each record is presented once, saving the storages and time. Parameters are initialized stochastically, and they are updated during the training process until termination. However, we omit this training of fake users' grading algorithm since its parameters are initialized already.

Let $result(u_j, i_k)$ be a record of recommendation in the training set where
\[
result(u_j, i_k) =
\begin{cases}
1, & \text{recommendation accepted}\\
-1, & \text{other }
\end{cases}.
\]
Consider the training error in the $n^{th}$ epoch
$$error(n) = result(u_j, i_k) - grade(u_j, i_k).$$
The gradients(partial derivatives) of $error(n)$ are
$$\frac{\partial}{\partial\alpha_1}error(n) = 2 fond(u_j, h_k) (hot_k - sim(u_j, i_k)),$$
$$\frac{\partial}{\partial\omega_1}error(n) = 2 \sigma \alpha_2 fond(u_j, h_k)\sum_{l = 1}^{m_j} \frac{\partial W_{jl}}{\partial\omega_1},$$
$$\frac{\partial}{\partial\omega_2}error(n) = 2 \sigma \alpha_2 fond(u_j, h_k)\sum_{l = 1}^{m_j} \frac{\partial W_{jl}}{\partial\omega_2},$$
where
$$\sigma = sgn(class\_weight(u_j) - class\_weight(i_k)),$$
$$W_{jl} = \overline{W}_{jl_1} + \widetilde{W}_{jl_2}.$$
$\overline{W}_{jl_1}$ is the weight of $u_j$'s own keyword class and $\widetilde{W}_{jl_2}$ is computed by searching related users(see section 3.3.3). By rounding $\frac{\partial \overline{W}_{kl_k}}{\partial\omega_i}fami(u_j, u_k)$:
\begin{equation*}
\begin{aligned}
\frac{\partial W_{jl}}{\partial\omega_1} &= \frac{\partial \widetilde{W}_{jl_2}}{\partial\omega_1}  \\
&\approx \frac{1}{2} \displaystyle\sum_{\substack{u_k \in related\_users(u_j), \\ class_{kl_k} = class_{jl_2}}} \overline{W}_{kl_k}(f(at) - f(comment)),
\end{aligned}
\end{equation*}
\begin{equation*}
\begin{aligned}
\frac{\partial W_{jl}}{\partial\omega_2} &= \frac{\partial \widetilde{W}_{jl_2}}{\partial\omega_2}   \\
&\approx \frac{1}{2} \displaystyle\sum_{\substack{u_k \in related\_users(u_j), \\ class_{kl_k} = class_{jl_2}}} \overline{W}_{kl_k}(f(retweet) - f(comment)).
\end{aligned}
\end{equation*}

After computation of the gradients we update $\alpha_1$ and $\omega_i(i = 1, 2)$ by applying momentum factor $\beta \in [0, 1]$(here the typical value $\beta = 0.9$). The parameter $\alpha_1$ in the $n^{th}$ epoch $\alpha_1(n)$ satisfies
$$\alpha_1(n + 1) = \alpha_1(n) + (1 - \beta)\frac{\partial}{\partial\alpha_1}error(n) + \beta \Delta \alpha_1(n),$$
where
$$\Delta \alpha_1(n) = \alpha_1(n) - \alpha_1(n-1).$$
$\omega_i, i = 1, 2$ are updated similarly. The training process is terminated in the $n^{th}$ epoch if no new training data $result(u_j, i_k)$ or $error(n) \le performance$, where $performance = 0.01$ controls the training process.

Furthermore, if we apply $revised\_grade()$ instead of $grade()$ in the grading process, the gradients should be revised as
$$\frac{\partial}{\partial\alpha_1}error(n) = \frac{2}{\lambda} time(u_j, h_k) fond(u_j, h_k) (hot_k - sim(u_j, i_k)),$$
$$\frac{\partial}{\partial\omega_1}error(n) = \frac{2}{\lambda} \sigma \alpha_2 time(u_j, h_k) fond(u_j, h_k)\sum_{l = 1}^{m_j} \frac{\partial W_{jl}}{\partial\omega_1},$$
$$\frac{\partial}{\partial\omega_2}error(n) = \frac{2}{\lambda} \sigma \alpha_2 time(u_j, h_k) fond(u_j, h_k)\sum_{l = 1}^{m_j} \frac{\partial W_{jl}}{\partial\omega_2}.$$

\section{Experiment and Improvement}
This section discusses the training result and evaluation metrics we adopted to test the system's performance. Some of other approaches are also introduced, which might enhance the performance of the recommender system.

\subsection{Training Result}
In our experiment we sampled 5,938 users' recommendation records from the dataset \cite{Niu:Tencent} stochastically and divided them into 2 subsets for training and testing. We omitted the update of $\omega_i$ to reduce the computational complexity and assigned $\omega_i = \frac{1}{3}$, which means at(@), retweet and comment occupy the same proportion when computing $fami(u_j, u_k)$.
We trained each user's patterns and computed the average parameters. Table \ref{train} presents the results of the training process.

\begin{table}[htbp]
\centering
\begin{tabular}{cccccc}
\hline
user class & user & followee & interaction & keyword & $\alpha_1$ \\
\hline
active & 3919 & 46 & 87 & 10 & 0.33 \\
inactive & 1194 & 27 & 42 & 8 & 0.18\\
fake & 825 & 18 & 2 & 5 & / \\
\hline
\end{tabular}
\caption{Training Sets and Optimal Parameters. Fake users' grading function has no parameters to update so we omit the training process of it.}
\label{train}
\end{table}

The result shows an evident discrepancy of $\alpha_1$, which reflects the inclination of accepting popular items. Inactive users prefer items with similar interests while active users prefer items with high popularity.

\subsection{Prediction and Precision Evaluation}
We computed $grade(u_j, i_k)$ of all $result(u_j, i_k)$ in testing subset and generated ordered item list of $u_j$(see section 3.3.4) to test the trained system. The evaluation metric is the average precision \cite{Zhu:AP} which KDD Cup's organizers adopted:
$$AP@3(u_j) = \sum_{i = 1}^{3} p(i)\Delta r(i),$$
where $p(i)$ is the precision of the $i^{th}$ recommended item and $\Delta r(i)$ is the change in the recall from $i - 1$ to $i$. Table \ref{Prediction} presents the $MAP@3$(mean value of $AP@3(u_j)$) results and Table \ref{Examples} presents the recommended item lists and the average precision of some users. The precision of fake users' prediction is much lower than others' in our experiment due to the difficulty of their interests' extractions. Adjusting $min\_action$ or recommending their linkers on other related platforms like QQ might help improve the results.
\begin{table}[!htb]
\centering
\begin{tabular}{cccc}
\hline
active & inactive & fake & total \\
\hline
0.41066 & 0.46879 & 0.33606 & 0.41198\\
\hline
\end{tabular}
\caption{Prediction Evaluation. Mining potential interests from inactive users' followees improves the performance of recommendation. Fake users' result is not good as the others.}
\label{Prediction}
\end{table}

\begin{table}[!htb]
\centering
\begin{tabular}{ccccc}
\hline
$u_j$ & user class & item & accepted item & $AP@3(u_j)$ \\
\hline
2071402 & active & 1606902 & 1606902 & 0.83 \\
        &        & 1760350 & 1774452 & \\
        &        & 1774452 &         & \\
\hline
942226 & inactive & 1606902 & 1606902 & 1.00 \\
       &          & 1606609 &         & \\
       &          & 1774452 &         & \\
\hline
193889 & fake & 1760642 & 1774862 & 0.33 \\
       &      & 1774684 &         & \\
       &      & 1774862 &         & \\
\hline
\end{tabular}
\caption{Examples of Prediction. User 2071402 accepts the $1^{st}$ and $3^{rd}$ items, then $AP@3 = (\frac{1}{1} + \frac{2}{3})/2 = \frac{5}{6}$; User 942226 only accepts the $1^{st}$ item, then $AP@3 = \frac{1}{1} = 1$; User 193889 only accepts the $3^{rd}$ item, then $AP@3 = \frac{1}{3}$.}
\label{Examples}
\end{table}

\subsection{Improvements of the System}
There are approaches to enhance the performance and overcome the limitations of our system.
Recommendation based on demographic methods \cite{DEMO} can help in enhancing the percentage of acceptance. Users with similar demographic information may have interest's coincidences thus accept similar items. 
Refined keyword analysis and user taxonomy can improve the recommendation. Users who follow items in the same category or interact with users who have explicit preferences can be grouped in identical user class. They share synonyms in their keywords and accept similar items in a high possibility based on the similarity of preferences.
Adaption to the frequently updated microblog platform's database can get user's present interests. Fortunately user's interests and behaviors are stable in a short period, so the system only needs retrains stochastically and gradually, which is fast and accurate.

\section{Conclusion and Future Work}
We present a hybrid recommender system for microblog to solve Track 1 task, KDD Cup 2012. The system analyzes the synonyms of keywords and behaviors of different users, extracts their (potential)interests, finds the target categories, grades the candidate items in those categories with indicators of popularity and similarity, and finally generates ordered item lists respect to each user. Experimental result shows high performance of our algorithm.
The initialization of grading function's parameters needs improvement, since good choices of them accelerate the process and avoid local minimums.
Dynamic algorithms which reduce the risk of inaccuracy by searching the best algorithm through competition also deserves further study.


\section{Acknowledgments}
We would like to thank the organizers of KDD Cup 2012 for organizing such a challenging and exciting competition. We would also like to thank Jiuya Wang for helpful discussions and Dachao Li from High- Performance Computing Platform of Sun Yat-Sen University for distributed computing supports.

%
\bibliographystyle{abbrv}
\bibliography{KDDCUPworkshop}  

\begin{thebibliography}{10}

\bibitem{KDD}
K.~C. 2012.
\newblock Predict which users (or information sources) one user might follow in
  tencent weibo.
\newblock http://www.kddcup2012.org/c/kddcup2012-track1, 2012.

\bibitem{Agrawal:AR}
R.~Agrawal, T.~Imieli\'{n}ski, and A.~Swami.
\newblock Mining association rules between sets of items in large databases.
\newblock In {\em Proceedings of the 1993 ACM SIGMOD international conference
  on Management of data}, SIGMOD '93, 1993.

\bibitem{Baike:Zombie}
Baidu.
\newblock Zombie fans on weibo.
\newblock http://baike.baidu.com/view/4047998.htm, 2010.

\bibitem{Han:FDM}
D.~Cheung, J.~Han, V.~Ng, A.~Fu, and Y.~Fu.
\newblock A fast distributed algorithm for mining association rules.
\newblock In {\em Parallel and Distributed Information Systems, 1996., Fourth
  International Conference on}, dec 1996.

\bibitem{Kon:apply}
J.~A. Konstan, B.~N. Miller, D.~Maltz, J.~L. Herlocker, L.~R. Gordon, and
  J.~Riedl.
\newblock Grouplens: applying collaborative filtering to usenet news.
\newblock {\em Commun. ACM}, 1997.

\bibitem{Kyle:Sina}
Kyle.
\newblock Sina commands 56\% of china's microblog market.
\newblock
  http://www.resonancechina.com/2011/03/30/sina-commands-56-of-chinas-microblog-market/,
  March 2011.

\bibitem{Mc:Birds}
M.~McPherson, L.~Smith-Lovin, and J.~M. Cook.
\newblock Birds of a feather: Homophily in social networks.
\newblock {\em Annual Review of Sociology}, 2001.

\bibitem{Niu:Tencent}
Y.~Niu, Y.~Wang, G.~Sun, A.~Y.~B. Dalessandro, C.~Perlich, and B.~Hamner.
\newblock {\em The Tencent Dataset and KDD-Cup'12}.
\newblock KDD-Cup Workshop, 2012.

\bibitem{Tan:DM}
P.~Tan, M.~Steinbach, and V.~Kumar.
\newblock {\em Introduction to Data Mining}, chapter 6. Association Analysis:
  Basic Concepts and Algorithms.
\newblock Addison-Wesley, 2005.

\bibitem{Tencent}
Tencent.
\newblock About tencent.
\newblock http://www.tencent.com/en-us/at/abouttencent.shtml, 2012.

\bibitem{QQ:Finance}
Tencent.
\newblock Tencent announces 2012 first quarter results.
\newblock
  http://www.tencent.com/en-us/content/ir/news/2012/attachments/20120516.pdf,
  May 2012.

\bibitem{Wang:Graph}
Z.~Wang, Y.~Tan, and M.~Zhang.
\newblock Graph-based recommendation on social networks.
\newblock In {\em Web Conference (APWEB), 2010 12th International
  Asia-Pacific}, 2010.

\bibitem{DEMO}
Wikipedia.
\newblock Demography.
\newblock http://en.wikipedia.org/wiki/Demography.

\bibitem{Zhu:AP}
M.~Zhu.
\newblock Recall, precision and average precision.
\newblock Working Paper 2004-09, Department of Statistics \& Actuarial Science,
  University of Waterloo, 2004.

\end{thebibliography}
%
%
\end{document}